\documentclass[a4paper,12pt]{article}

\usepackage[english]{babel}
\usepackage{times}
\usepackage{latexsym} 
\usepackage{amsmath}
\usepackage{amssymb}
\usepackage{graphicx}
\usepackage{wrapfig}
\usepackage{indentfirst}

\makeatletter
\def\@seccntformat#1{\csname the#1\endcsname.\quad}
\makeatother

\newcommand{\be}{\begin{equation}}
\newcommand{\ee}{\end{equation}}
\newcommand{\ba}{\begin{eqnarray}}
\newcommand{\ea}{\end{eqnarray}}
\let\f\frac

\newcommand{\ds}{\displaystyle}

\begin{document}

\title{\LARGE\textbf{On one fundamental property\\
of gravitational field in the field theory}}
\date{}
\author{\normalsize S.\,S. Gershtein\footnote{gershtein@mx.ihep.su}, A.\,A. Logunov\footnote{Anatoly.Logunov@ihep.ru},
        M.\,A. Mestvirishvili,\\
{\small{\it Institute for High Energy Physics, Protvino, Russia}}} \maketitle

\begin{abstract}\noindent
{It is shown that the universal property of  gravitational field
to slow down the rate of time leads in the field theory to a
fundamental property --- generation of effective forces of
repulsion.}
\end{abstract}

There is a common belief that the gravitational field provides only forces of attraction. It is seen, for example,
from the fact that the physical velocity of a test body increases when it is approaching the gravitating body.
However this is not quite so for strong fields. Let us consider this later. When A.\, Einstein  had connected
gravitational field with the Riemannian space metric tensor in 1912 it was found that such a field was slowing
down the rate of time for a physical process. This slowing down can be demonstrated, in particular, by the example
of Schwarzschild solution, if we compare the rate of time in the gravitational field with the rate of time for a
distant observer. Nevertheless in  general case there is only Riemannian space metric tensor in General Relativity
and there are not any indication of the inertial time of Minkowski space. Just for this reason the universal
property of gravitational field to slow down the time rate in comparison to the inertial time could not be further
developed in General Relativity. The situation is rather opposite in the Relativistic  Theory of Gravitation (RTG)
as it is a field theory. In this approach the gravitational field is treated as a physical field of
Faraday-Maxwell type which is developing in Minkowski space in line with all the other physical fields.

The source of universal gravitational field is the total conserved energy-mo\-men\-tum tensor of all matter
including the gravitational field. Therefore the gravitational field is a tensor field with spins $2$ and $0$.
Just this fact leads to \textbf{geometrization}: the effective Riemannian space arises but with trivial topology.
This leads to the following situation: the motion of a test body in Minkowski space under the action of
gravitational field is equivalent to the motion of this body in the effective Riemannian space created by this
gravitational field. The arising of the effective Riemannian space in the field theory side by side with
preserving the role of Minkowski space as the fundamental space gives a special meaning to the property of the
gravitational field to slow down the time rate. Just in this case it is only possible to speak about the slowing
of time in full, by comparing the time rate in the gravitational field with the time rate in inertial frame of
reference of Minkowski space in the absence of gravitation. And all this is realized in the RTG because the metric
tensor of Minkowski space enters into the full system of its equations. But this general property of the
gravitational field to slow down the time rate leads in the field theory to a remarkable conclusion: \textbf{the
slowing down of the physical process time rate in a strong field generates effective field forces of the
gravitational nature. These effective forces in gravitation occur to be repulsive}.

 To
demonstrate that a change of the time rate leads to arising of a
force let us consider Newton equation
\[
\f{d^2x}{dt^2}=F\,.
\]
If we formally transform this equation in order to change the
inertial time $t$ for time $\tau$ according to the rule
\[
d\tau =U(t)dt\,,
\]
then we easily get
\[
\f{d^2x}{d\tau^2}=\f{1}{U^2}\Bigl \{F-\f{dx}{dt}\f{d}{dt}\ln U
\Bigr \}\,.
\]
It is seen from here that a change in time rate determined by
function $U$ leads to arising of the effective force. But all this
is of purely formal character because there are no any physical
reason in this case that could change the time rate. But just this
formal example demonstrates that when a process of slowing down
time takes place in  nature it inevitably generates effective
field forces, and therefore it is necessary to account for them in
the theory as something rather new and surprising.

And just  here  we are to turn to gravitation. The physical
gravitational field changes both the time rate and parameters of
spatial quantities in comparison to their values given in
 inertial system of Minkowski space when gravitation is absent. Just
for this reason all this should be taken into account in the
gravitational field equations. In the RTG metric tensor of
Minkowski space appears unambiguously in these equations due to
graviton mass introduced into them. Just this tensor provides the
opportunity to account for effective field forces created by
change of the time rate under action of the gravitational field.
Here the graviton mass realizes a correspondence of the effective
Riemannian space to the basic Minkowski space. Though the graviton
mass is rather small nevertheless the influence of mass term
becomes decisive because of great slowing down of the time rate
under the action of gravitational field.

The complete system of RTG equations can be displayed as follows [1, 2]: \be
\Bigl(R^{\mu\nu}-\f{\,1\,}{2}g^{\mu\nu}R\Bigr)+\f{m_g^2}{2} \Bigl[g^{\mu\nu}+\Bigl(g^{\mu\alpha}g^{\nu\beta}
-\f{\,1\,}{2}g^{\mu\nu}g^{\alpha\beta}\Bigr)\gamma_{\alpha\beta}\Bigr] =8\pi GT^{\mu\nu}\,, \label{eq1} \ee \be
D_\nu\tilde{g}^{\nu\mu}=0\,. \label{eq2} \ee Here $D_\nu$ is the covariant derivative in Minkowski space,
$\gamma_{\alpha\beta}$ ---  metric tensor of Minkowski space, $g_{\alpha\beta}$
---  metric tensor of the effective Riemannian space, $m_g$ --- graviton mass. This system of equations is
generally covariant under arbitrary coordinate transformations and
form-invariant under Lorentz trans\-for\-ma\-ti\-ons.

Now let us demonstrate by examples of the collapse and evolution
of the homogeneous isotropic Universe how effective field
repulsive forces are developed  due to slowing down of the time
rate under action of the gravitational field. Consider the static
spherically symmetric field \be ds^2 =U(r)dt^2 -V(r)dr^2-W^2(r)
(d\varTheta^2+\sin^2\varTheta\,d\phi^2)\,, \label{eq3} \ee \be
d\sigma^2 =dt^2 -dr^2-r^2
(d\varTheta^2+\sin^2\varTheta\,d\phi^2)\,. \label{eq4} \ee Here
function $U(r)$ determines slowing down of the time rate in
comparison to inertial time $t$.

Strong slowing down of the time rate takes place when this
function is small enough in comparison to unity.  Eqs. (\ref{eq1})
take the following form for this problem defined by (\ref{eq3})
and (\ref{eq4}):
\ba \label{eq5} &L_1=8\pi G\rho
-\ds\f{\,1\,}{2}m_g^2
\Bigl[1+\f{\,1\,}{2}\Bigl(\f{\,1\,}{U}-\f{\,1\,}{V}\Bigr)
-\ds\f{r^2}{W^2} \Bigr]\nonumber\\[2mm]
&L_2=-8\pi Gp -\ds\f{\,1\,}{2}m_g^2
\Bigl[1-\f{\,1\,}{2}\Bigl(\f{\,1\,}{U}-\f{\,1\,}{V}\Bigr)
-\f{r^2}{W^2} \Bigr]\\[2mm]
&L_3=-8\pi Gp -\f{\,1\,}{2}m_g^2 \Bigl[1-\ds\f{\,1\,}{2}\Bigl(\f{\,1\,}{U}+\f{\,1\,}{V}\Bigr) \Bigr]\nonumber \ea
If we put the graviton mass $m_g$  equal to zero, then system of equations (\ref{eq5}) will coincide with the
Hilbert-Einstein system of equations for problem given by Eqs. (\ref{eq3}) and (\ref{eq4}). In this case it will
have the celebrated Schwarzschild solution, when functions $U$, $V$ and $W$
 given as follows \be \label{eq6} U=\f{r-GM}{r+GM},\quad
V=\f{r+GM}{r-GM},\quad W=(r+GM)\,. \ee It could be seen from here,
in particular, that strong slowing down of the time rate in
comparison to inertial time $t$ takes place in the region where
$r$ is close to $GM$. But due to smallness of $U$ the terms
standing in r.h.s. of Eq. (\ref{eq5}) and having $U$ in their
denominator will be dominate. Just this fact leads, after careful
analysis  [1--3] accounting for the graviton mass, to new
expressions for functions $U$ and $V$ which are rather different
from (\ref{eq6}): \be \label{eq7} U=\Bigl(\f{GM m_g}{\hbar
c}\Bigr)^2,\quad V=\f{\,1\,}{2}\f{W}{W-2GM}\,. \ee This provides
the stopping of test body motion creating a turning point where
velocity $dW/ds=0$ \be \label{eq8} \f{dW}{ds}=-\f{\hbar c^2}{m_g
GM}\Bigl[\f{W}{GM} \Bigl(1-\f{2GM}{c^2W}\Bigr)\Bigr]^{1/2}\,. \ee
The acceleration at this turning point is as follows \be
\label{eq9} \f{d^2W}{ds^2}=\f{1}{2GM}\Bigl(\f{\hbar c^2}{m_g
GM}\Bigr)^2\,,\ee it is positive and it corresponds to a repulsive
force. Just for this reason Schwarzschild singularity  and
therefore also the possibility of ``black holes'' formation is
excluded.

Another example, demonstrating appearance of new effective field
forces due to slowing down of the time rate is the evolution of
homogeneous isotropic Universe. In this case we obtain, on the
base of Eq. (\ref{eq2}), the flat Universe solution with only
Euclidean 3-dimensional geometry, i.e. \ba &ds^2 =d\tau^2 -\beta^4
a^2
(dr^2+r^2d\varTheta^2+r^2\sin^2\varTheta\,d\phi^2)\,,\nonumber\\
\label{eq10}\\
&d\sigma^2 =\ds\f{1}{a^6}d\tau^2
-dr^2-r^2d\varTheta^2-r^2\sin^2\varTheta\,d\phi^2\,.\nonumber \ea
Here $\beta^4$ is a constant determined by the integral of motion.
The proper time $d\tau$ is related to inertial time $dt$ by the
following formula \be \label{eq11} d\tau =a^3dt \ee

Eqs. (\ref{eq1}) can be reduced, on the base of Eqs. (\ref{eq10}),
to the following system of equations for  scale factor  $a(\tau)$
[1, 2, 4]: \be \label{eq12}
\Bigl(\f{\,1\,}{a}\f{da}{d\tau}\Bigr)^2 =\f{8\pi G}{3}\rho (\tau)
-\f{1}{12}m_g^2 \Bigl(2-\f{3}{\beta^4 a^2}+\f{1}{a^6}\Bigr)\,, \ee
\be \label{eq13} \f{\,1\,}{a}\f{d^2a}{d\tau^2} =-\f{4\pi
G}{3}\Bigl(\rho +\f{3p}{c^2}\Bigr) -\f{\,1\,}{6}m_g^2
\Bigl(1-\f{1}{a^6}\Bigr)\,, \ee 
here we assume $\hbar=c=1$. The scale factor $a(\tau)$
determines, through Eq. (\ref{eq11}), the slowing down of the time
rate produced by the action of gravitational field. But just this
factor, standing in r.h.s. of Eq. (\ref{eq12}) stops the process
of Universe contraction when strong slowing down of the time rate.
Then the minimal value of $a$ is as follows
\[
 a_{\min}
 =\Bigl[\Bigl(\f{m_g c^2}{\hbar}\Bigr)^2
 \f{1}{32\pi G\rho_{\max}}\Bigr]^{1/6}\,,
 \]
so the cosmological singularity is removed. From the other side,
the repulsive force, created due to the slowing down of time rate,
provides accelerated expansion of the Universe as follows from
Eq.~(\ref{eq13}).

At the point of stopping the contraction acceleration is given as
follows
\[
\f{\,1\,}{a}\ds\f{d^2 a}{d\tau^2}\bigg|_{\tau =\,0} =\f{8\pi
G}{3}\rho_{\max}\,.
\]
Just this was ``the impulse'' to begin the Universe expansion. So,
the field ideas of the gravitational field as a physical one give
us the opportunity \textbf{to discover a fundamental property of
the gravitational field --- to create effective repulsive forces
in strong fields due to slowing down the time rate}. There are no
such forces in General Relativity.
An interesting pattern arises: in RTG the gravitational field,
developing itself through attraction forces and collecting matter,
then enters a stage when, under the action of this strong field,
the slowing down of time rate in comparison to the inertial time
is starting, and this inevitably leads to creation of effective
repulsive field forces which stop the process of matter
contraction under attraction forces, providing later a process of
expansion. We see that a special mechanism of self-regulation is
supplied to the gravitational field in the field theory. Just this
realizes  stopping of the massive bodies collapse and removes the
cosmological singularity, providing cyclic development of the
Universe.

The authors express their deep gratitude to V.\,I.~Denisov, V.\,A.~Petrov, N.\,E.~Tyu\-rin and Yu.\,V.~Chugreev
for valuable discussions.

\end{document}